\documentclass[Journal,10pt]{IEEEtran}

\usepackage{color}
\usepackage{float}
\usepackage{tabularx,colortbl}
\usepackage{setspace}
\usepackage{cite,graphicx,amsmath,psfrag,bm}
\usepackage{subfigure}
\usepackage[usenames,dvipsnames]{xcolor}
\usepackage{mathabx}
\usepackage{cancel}
\usepackage{epstopdf}
\usepackage{amsmath}
\usepackage{cancel}

\usepackage[process=auto]{pstool}
\usepackage[normalem]{ulem}
\graphicspath{{Figs/}}

\newcommand{\figref}{Fig.~\ref}

\ifCLASSINFOpdf
\else
\fi

\begin{document}

\title{Real-Time Spectrum Sniffer for Cognitive Radio \\ Based on Rotman Lens Spectrum Decomposer}

\author{Xiaoyi~Wang,~\IEEEmembership{Student Member,~IEEE,}
        Alireza~Akbarzadeh,
        Lianfeng~Zou,
        and~Christophe~Caloz,~\IEEEmembership{Fellow,~IEEE}
\thanks{
The authors are with the department
of electrical engineering, Polytechnique Montr\'{e}al,
Montr\'{e}al, Qu\'{e}bec, Canada. (E-mail: xiaoyi.wang@polymtl.ca).}
}

\maketitle
\begin{abstract}
We introduce the concept of a Rotman-lens spectrum decomposer (RLSD) real-time spectrum-sniffer (RTSS) for cognitive radio. Compared to a previously existing RTSS, the RLSD-RTSS offers the advantages of being 1)~based an a simpler and lower-cost purely passive structure, 2)~easier to design and easily amenable to tunability, 3)~of much broader bandwidth, and 4)~of accommodating more channels. The electrical size of the device is electrically larger, but perfectly acceptable in the millimeter-wave frequency range. The proposed RLSD-RTSS is demonstrated theoretically and experimentally, and been shown to support tunability in terms of both bandwidth-resolution and operation band. Given its unique features, this device may find wide applications in 5G UHD and 3D video systems.
\end{abstract}


\section{Introduction}

Current wireless communication systems will need drastic improvement to accommodate future end-user speed and reliability requirements. Real-time reconfiguration strategies, globally referred to as \emph{cognitive radio}~\cite{Jour:Liang_2011_TVT_CognitiveRadio}, may become pivotal approach to address this challenge in forthcoming 5G systems, especially in applications with extremely high throughput, such as UHD and 3D video~\cite{Book:Osseiran_2016_5G}.

Cognitive radio consists in two main steps: 1)~sensing -- or ``\emph{sniffing}'' -- the ambient spectrum so as to identity white (free) bands in it, and 2)~reconfigure the radio system to exploit these white bands for optimal spectral efficiency at all times. As the operating frequency gets higher, the sniffing operation should be realized in a fast, adaptive and low-cost fashion. \emph{Real-time analog processing~(RAP)}~\cite{Jour:2013_MwMag_Caloz,Jour:Zou_2017_TWC_DCMA,JOUR:2003_TMTT_Laso,JOUR:2015_Gomez-Tornero_TMTT_SIW_Multiplexer,JOUR:2009_Gupta_TMTT_RTSA,JOUR:2014_MWCL_Nikfal,Conf:Wang_AP-S2017_Phaser}, based on agile microwave dispersive components called ``phasers'' \cite{JOUR:2011_TMTT_Nikfal,JOUR:2012_MWCL_Horii,JOUR:2012_TMTT_Zhang,JOUR:2012_TMTT_Shulabh,JOUR:2014_JRMCAE_Zhang,JOUR:2015_TMTT_Gupta,Conf:Wang_URSIGASS2017}, is an optimal technology in this regard.

Recently, a Real-Time Spectrum Sniffer (RTSS) based on a mixer and a coupled-line phaser with stair-case group delay response has been reported in~\cite{Jour:Nikfal_2012_MWCL_SpectrumSniffer}. However, this sniffer is limited by the following features: requirement for an auxiliary pulse generator and of a local oscillator source for mixing, high design complexity and lack of tunability, restricted bandwidth, and small number of channels.

In this paper, we present an alternative RTSS, based on a Rotman Lens Spectrum Decomposer (RLSD)~\cite{Conf:Fusco_Multiplerxer_2012,Jour:Wang_2017_RLSD}. This device resolves all the aforementioned issues of the RTSS in~\cite{Jour:Nikfal_2012_MWCL_SpectrumSniffer}, and is hence very promising for future communication systems: 1)~it is based on a simple passive structure, the RLSD, requiring neither mixers nor sources, and it is hence inexpensive; 2)~it easy to design and may be tuned in real-time using PIN diodes and switches; 3)~it exhibits a very broad bandwidth, due to its true-time delay nature; 4)~it may accommodate a great number channels.

The paper is organized as follows. Section~\ref{SEC:RTSS} presents the principle of the proposed RTSS. Section~\ref{SEC:RLSD} recalls the fundamental aspects of the RLSD that constitutes the key component of the RTSS. Section~\ref{SEC:Res_Control} shows the resolution control capabilities of the RLSD-RTSS, which include resolution uniformity, bandwidth resolution trade-off tuning and operation band tuning. The RLSD-RTSS is demonstrated in Sec.~\ref{SEC:Demonstration}, and Sec.~\ref{SEC:Conclusion} concludes the paper.

\section{Real-time Spectrum Sniffer (RTSS) Principle}\label{SEC:RTSS}

\begin{figure}[h!t]
    \centering
    \includegraphics[width=\columnwidth]{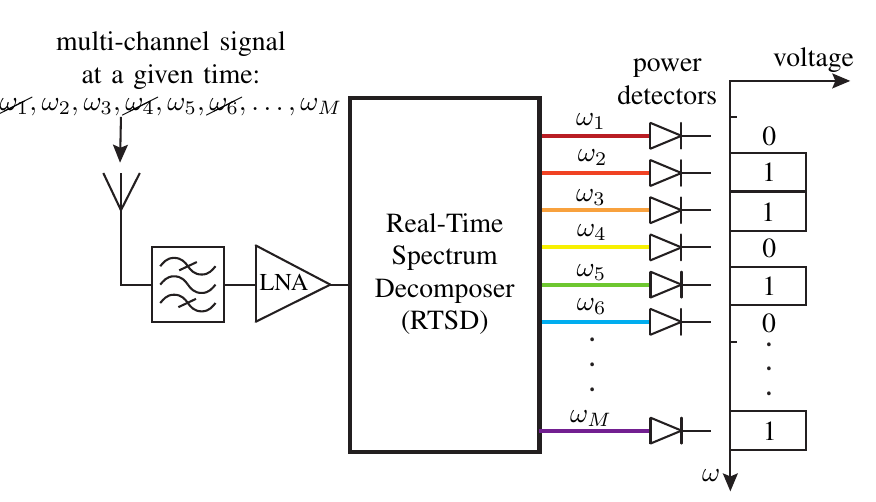}
        \caption{Principle of the proposed real-time spectrum sniffer (RTSS).}
   \label{FIG:principle}
\end{figure}

The principle of the proposed real-time spectrum sniffer (RTSS) is presented in \figref{FIG:principle}. The broadband-spectrum (multi-channel) ambient signal to sniff is picked up by an omnidirectional antenna, amplified and passed through a real-time spectrum decomposer (RTSD). The RTSD operates like a prism, i.e. separates out in space, towards different output ports, (and in real-time) the different frequencies composing the input signal and corresponding to different communication channels. The so-separated output signals are then detected by power detectors, from which the binary information on the activity (bit 1) or inactivity (bit 0) of all the channels is instantaneously provided in the base-band domain.

\section{Rotman Lens Spectrum Decomposer (RLSD)} \label{SEC:RLSD}

\begin{figure}[h!t]
    \centering
    \includegraphics[width=\columnwidth]{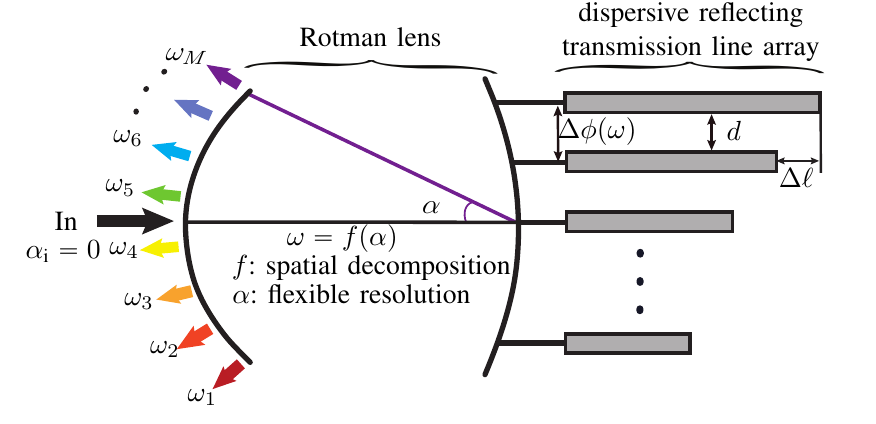}
        \caption{Implementation of the real-time spectrum decomposer (RTSD) in~\figref{FIG:principle} based on a Rotman lens and a dispersive reflecting transmission line array (lens RTSD).}
   \label{FIG:RTSD}
\end{figure}

The key component in the RTSS of \figref{FIG:principle} is the real-time spectrum decomposer (RTSD). We propose here to realize this RTSD in the form of the ray-optics structure shown in \figref{FIG:RTSD}. This structure is composed by the interconnection of a Rotman lens~\cite{JOUR:Rotman_Rotmanlens_1963,JOUR:Hansen_Rotmanlenses_1991,Jour:Rotman_ProcIEEE_TrueTimeDelay,JOUR:2014_CJE_Vashist_ReviewRotamLens,Thesis:Dong_RotmanLens_2009} and a reflective dispersive transmission line array~\cite{Jour:1996_Smit_AWG}.

The basic operation of the lens RTSD is as follows. The ambient broadband input signal is injected into the system from a port at the left center of the lens and then cylindrically radiates within the lens to the ports at its right. The (full-spectrum) signals reaching these ports are then reflected by an array of transmission lines with different lengths. Due to this dispersive nature of this array reflector~\cite{Jour:1996_Smit_AWG}, a phase gradient is formed on the right ports, causing the wave to decompose towards the left, with different frequencies reaching different ports, so as to achieve the desired spectral decomposition.
\vspace{-0.1cm}

Calling $\Delta\ell$ the length difference between adjacent transmission lines in~\figref{FIG:RTSD}, the phase gradient after round-trip reflection from the dispersive transmission line array is found as
\begin{equation}\label{EQ:Phi}
\Delta\phi_\text{array}(\omega)=2\beta_\text{e}(\omega)\Delta\ell=\frac{2\omega\sqrt{\epsilon_\text{e}}}{c}\Delta\ell,
\end{equation}
where $\beta_\text{e}(\omega)$ is the effective wavenumber of the transmission lines. Assuming that $\Delta\ell$ is designed to be $N$ times half the wavelength at the frequency intended to scatter at the center of the RLSD left contour, $\omega_0$, i.e.
\begin{equation}\label{EQ:DeltaL}
\Delta\ell
=N\lambda_0/2
=N\pi c/(\sqrt{\epsilon_\text{e}}\omega_0),\quad N=1,2,3,\ldots,
\end{equation}
we find, upon substitution into~\eqref{EQ:Phi},
\begin{equation}\label{EQ:Phi_Array}
\Delta\phi_\text{array}(\omega)=2\pi N\frac{\omega}{\omega_0}.
\end{equation}

On the other hand, according to antenna array theory, the function $\Delta\phi_\text{lens}(\omega)$ for beam forming at the angle $\psi$ is
\begin{equation}\label{EQ:Phi_Lens}
\Delta\phi_\text{lens}(\omega)=\frac{\omega}{c}d\sin{\psi}=\frac{\omega}{c}d\gamma\sin{\alpha},
\end{equation}
where the latter equality stems from the Rotman lens beam former relationship~\cite{JOUR:Rotman_Rotmanlens_1963}
\begin{equation}\label{EQ:gamma}
\gamma\approx\frac{\sin{\psi}}{\sin{\alpha}},
\end{equation}
with $\alpha$ being the port position angle, as shown in \figref{FIG:RTSD}.

$\Delta\phi_\text{lens}(\omega)$ represents the phase gradient at the right contour of the Rotman lens that is required for the frequency $\omega$ to radiate into the direction $\alpha$, as required for spectral decomposition. This gradient must be compatible with the phase gradient after reflection from the dispersive array, $\Delta\phi_\text{array}$. For this to be the case, we equate~\eqref{EQ:Phi_Array} and~\eqref{EQ:Phi_Lens}, which yields the frequency ($\omega$) versus port position ($\alpha$) law of the lens RTSD:
\begin{equation}\label{EQ:Omega_Alpha}
\omega(\alpha)=\frac{2\pi N\omega_0c}{2\pi Nc-\omega_0d\gamma\sin{\alpha}}.
\end{equation}

\section{RTSD Resolution Control}\label{SEC:Res_Control}

\subsection{Uniform Resolution Design}\label{SEC:Unif_res_des}
If the output ports of the lens RTSD are equidistant, i.e. $\Delta\alpha=\alpha_{k+1}-\alpha_k=\text{constant}$ in \figref{FIG:RTSD}, the resolution across the spectrum of the input signal is nonuniform~\cite{Conf:Fusco_Multiplerxer_2012}, which is most often not desired in practice. Therefore, we shall distribute the output ports in such a fashion that the RTSD exhibits \emph{uniform resolution}~\cite{Jour:Wang_2017_RLSD}.

\begin{figure}[h!t]
    \centering
     \includegraphics[width=0.95\columnwidth]{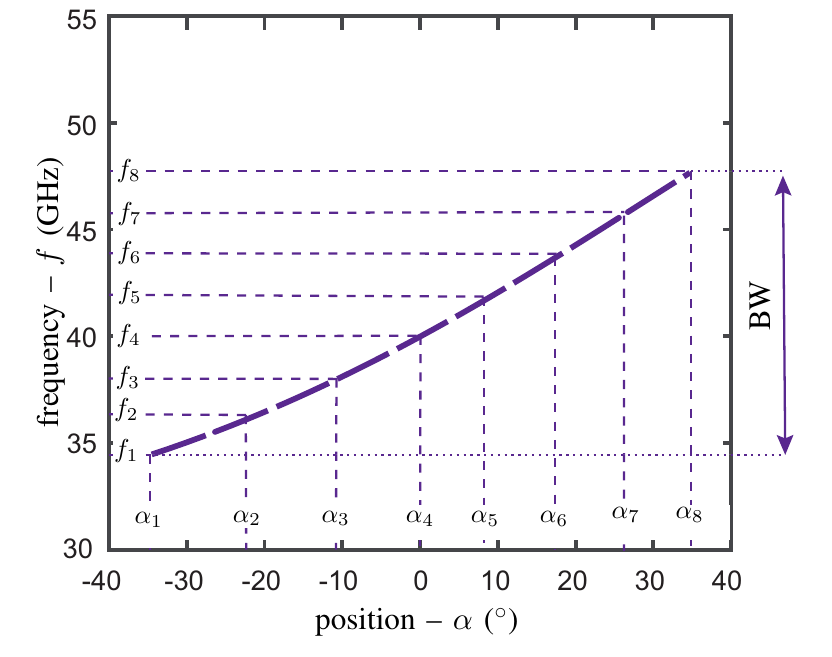}
        \caption{Frequency ($f$) versus port position ($\alpha$) of the lens RTSD in \figref{FIG:RTSD}, obtained by~\eqref{EQ:Omega_Alpha}, with $f_0=\omega_0/(2\pi)=40$  GHz, $N=2$ and $\gamma=1$. The discrete port positions for \emph{uniform-resolution} [$\Delta f_k=f_{k+1}-f_k=\Delta f=\text{const.}$ ($k=1,\ldots,7$)] are shown for $N=2$.}
   \label{FIG:Omega_Alpha}
\end{figure}

For the sake of concreteness, let us develop the overall RTSD resolution control theory through a specific design example. Consider an RTSS operation around $f_0=\omega_0/(2\pi)=40$~GHz with bandwidth resolution of about 15~GHz,  and $M=8$~output ports (\figref{FIG:principle}), with the parameters $N=2$, $d=\lambda_0/2$ or $d=c/(2f_0)$ and $\gamma=1$. Inserting these parameters in~\eqref{EQ:Omega_Alpha} leads to the frequency versus position relation plotted in \figref{FIG:Omega_Alpha}.

Achieving resolution uniformity simply consists in uniformly sampling the curve $f(\alpha)$ in \figref{FIG:Omega_Alpha} along the vertical (frequency--$f$) axis and reading out the corresponding angular positions on the horizontal (position--$\alpha$) axis. The values $\alpha_n$ obtained in this fashion directly provide the locations of the ports on the left contour of the lens RTSD producing uniform resolution.

\begin{figure}[h!t]
    \centering
    \includegraphics[width=0.95\columnwidth]{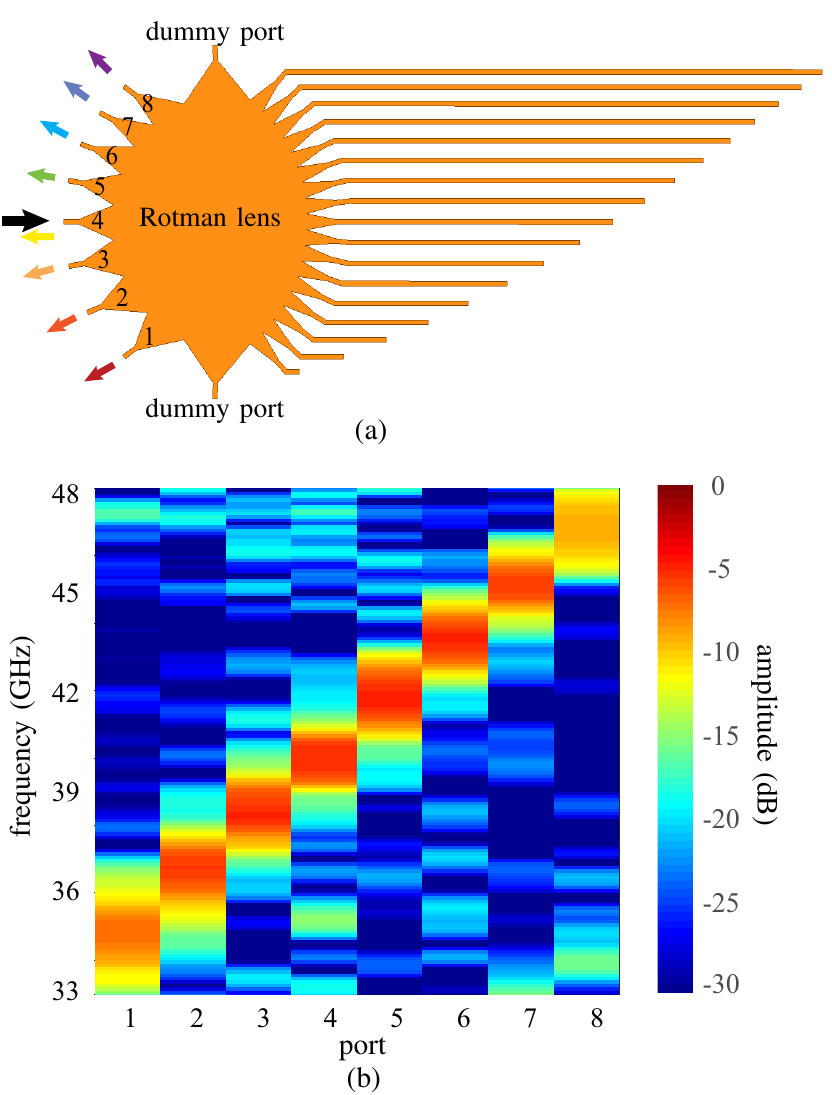}
        \caption{Uniform-resolution 8-channel RTSD lens. (a)~Layout. (b)~Full-wave simulated spectrum versus lens port.}
   \label{FIG:Layout}
\end{figure}

The 8-channel uniform-resolution lens RTSD is then designed in microstrip technology on a Rogers 6002 substrate with thickness 0.254~mm, dielectric constant~2.94 and loss tangent~0.0012. The resulting layout is shown in Fig.~\ref{FIG:Layout}(a). The Rotman lens includes 8 input ports, 15 output ports and 2 dummy ports (reducing spurious reflection~\cite{JOUR:Rotman_Rotmanlens_1963}). The 3 focal angles~\cite{JOUR:Rotman_Rotmanlens_1963} are $0^\circ$, $\alpha_0=35^\circ$ and $-\alpha_0=-35^\circ$, with focal length $28.4$~mm, $25.6$~mm and $25.6$~mm, respectively.
The 15 transmission lines are directly connected to the output ports of the Rotman lens with adjacent line length difference of 4.8~mm and uniform width of 0.66~mm .

Figure~\ref{FIG:Layout}(b) plots the full-wave simulated spectrum versus port of the RTSD lens in \figref{FIG:Layout}(a). It shows that the lens RTSD essentially operates as expected, despite some spurious inter-port leakage due to small mismatch at the lens to strip transitions and resulting inter-port coupling. This effect may be mitigated by using longer or more efficient (e.g. exponential or Klopfenstein~\cite{BK:2011_Pozar}) transitions.

\subsection{Bandwidth-Resolution Tuning}
Tuning the length difference $\Delta\ell$ of the reflecting array transmission lines [Eq.~\eqref{EQ:Phi}], or equivalently tuning $N$ since $\Delta\ell~\propto~N$ [Eq.~\eqref{EQ:DeltaL}], tunes the phase gradient $\Delta\phi_\text{array}$ of the array [Eq.~\eqref{EQ:Phi_Array}], which results in altering the receive angle versus frequency sensitivity, $s(N)=\partial\alpha/\partial\omega$ and hence the bandwidth of the RTSS. Specifically, increasing $N$ increases the array dispersion and lens diffraction, i.e. splits frequencies more over space or increases $s(N)$, which decreases the RTSS bandwidth, BW$=[\omega(-\alpha_0),\omega(\alpha_0)]$, since the extremal frequencies of the test pulse are pushed beyond the receiving area ($[-\alpha_0,\alpha_0]$) of the lens. So, we should have $\text{BW}~\propto~1/s~\propto~\partial\omega/\partial\alpha~\propto~1/N$.

Let us verify this mathematically by deriving~\eqref{EQ:Omega_Alpha}
\begin{equation}\label{EQ:Sensitivity I}
\text{BW}(N)\propto~\frac{\partial\omega}{\partial\alpha}=\frac{2\pi N\omega^2d\gamma\cos{\alpha}}{(2\pi Nc-\omega_0d\gamma\sin{\alpha})^2}.
\end{equation}
This expression is rather complicated, but its evaluation about the center port of the RTSS, $\alpha=0$, leads to the much simpler result
\begin{equation}\label{EQ:Sensitivity_IE}
\text{BW}(N)~\propto~\left.\frac{\partial\omega}{\partial\alpha}\right|_{\alpha=0}=\frac{\omega_0}{2N},
\end{equation}
which confirms the predicted trend of the bandwidth being inversely proportional to $N$.

\begin{figure}[h!t]
    \centering
    \includegraphics[width=\columnwidth]{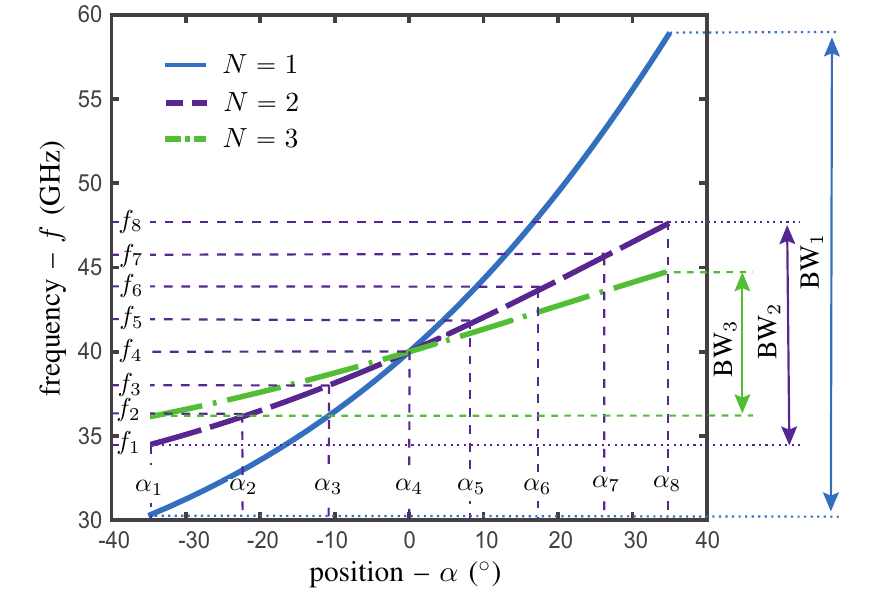}
        \caption{Frequency ($f$) versus port position ($\alpha$) of the lens RTSD in \figref{FIG:RTSD}, computed by~\eqref{EQ:Omega_Alpha} for $N=1,2,3$ in~\eqref{EQ:DeltaL} with $f_0=\omega_0/(2\pi)=40$ GHz. The bandwidths for $N=1,2,3$ are [30.5,58.5]~GHz, [35.0,46.7]~GHz and [36.2, 43.5]~GHz, respectively.
        }
   \label{FIG:Omega_Alpha_N}
\end{figure}

Figure~\ref{FIG:Omega_Alpha_N} plots the function $\omega(\alpha)$ given by~\eqref{EQ:Omega_Alpha} for $N=1,2,3$, where $N=2$ corresponds to the design in \figref{FIG:Omega_Alpha}. By design [from~\eqref{EQ:Phi_Array} $\Delta\phi_\text{array}(\omega=\omega_0)=2\pi N$, yielding the same radiation pattern, with main beam pointing to $\alpha=0$, $\forall~N$], the three curves intersect at the point $(0,f_0)=(0^\circ,40~\text{GHz})$, and
the operating frequency range indeed increases with decreasing~$N$. The operation bandwidth for $N=1$ is almost five times larger than that for $N=3$, and this ratio would naturally increases with larger $N$. This represents a very large bandwidth tuning capability, where one can trade bandwidth for resolution and vice-versa.

\begin{figure}[h!t]
    \centering
    \includegraphics[width=0.8\columnwidth]{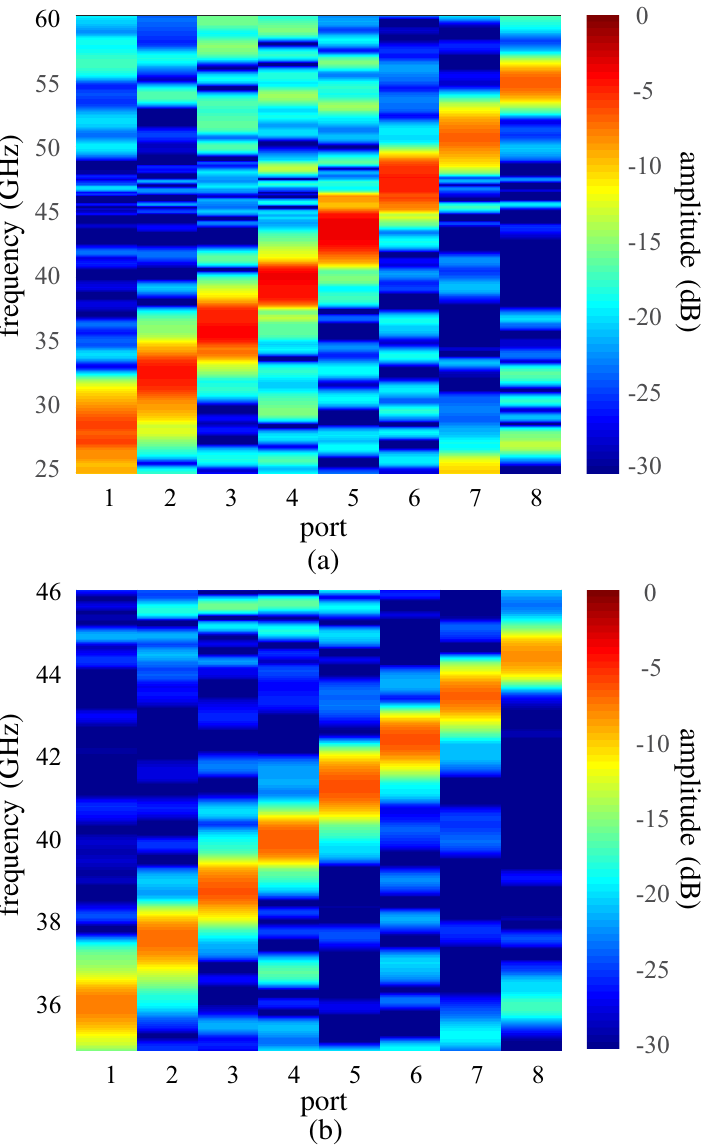}
        \caption{Full-wave simulated spectra versus lens port number in the RTSD of~\figref{FIG:Layout}(a) for different values of $N$. (a)~$N=1$. (b)~$N=3$. The result for $N=2$ is plotted in \figref{FIG:Layout}(a).}
   \label{FIG:S_N}
\end{figure}

Figure~\ref{FIG:S_N} plots the full-wave simulated spectra versus lens port number in the RTSD of \figref{FIG:Layout}(a) for $N=1$ and $N=3$, complementing the results for $N=2$ in \figref{FIG:Layout}(b). One first observes that the operation bandwidths correspond to the theoretical predictions in \figref{FIG:Omega_Alpha_N}. The second observation is that the amount of parasitic leakage increases with decreasing $N$. This is simply understood by remembering that the bandwidth is inversely proportional to $N$ [Eq.~\eqref{EQ:Sensitivity_IE}] and realizing the limited bandwidth of the transitions between the lens and the microstrip lines on both sides of the lens.

\begin{figure}[h!t]
    \centering
    \includegraphics[width=\columnwidth]{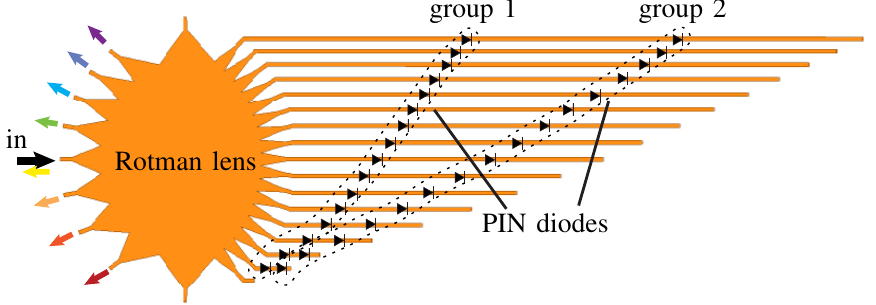}
        \caption{Implementation of a 3-mode ($N=1,2,3$ in \figref{FIG:Omega_Alpha_N}) bandwidth-resolution tuning by RLSD-RTSS using 2 PIN diodes in each reflecting line. In the first mode, the diodes of group~1 are off ($N=1$ or $\Delta\ell=\lambda_0/2$); in the second mode, these diodes are on and those of group 2 are off ($N=2$ or $\Delta\ell=\lambda_0$); in the third mode, all the diodes are on ($N=3$ or $\Delta\ell=3\lambda_0/2$).}
   \label{FIG:Implement_1}
\end{figure}

Practically, the bandwidth-resolution tuning demonstrated in this section can be straightforwardly implemented by multiplying all the lengths of the reflecting array transmission lines by $N$ and connecting the resulting $N$ sections by PIN diodes, as is shown in~\figref{FIG:Implement_1}. If all the diodes are ON, then the RTSS exhibits the highest resolution, while if the first diode towards the right is OFF, the RTSS provides the largest bandwidth.

\subsection{Operation Band Tuning}
Until this point, the input port has always been fixed to the middle of the left contour of the Rotman lens, i.e. $\alpha_\text{i}=0$, as shown in \figref{FIG:RTSD}. Switching the input position between the ports the RTSS, i.e. letting $\alpha_\text{i}=\alpha_k$ with $k$ \emph{varying} between $k=1$ to $k=M$, slides the operating spectrum and hence allows sniffing different frequency bands. Indeed, such switching changes the optical path lengths across the structure, effectively adding an extra term to the array phase gradient $\Delta\phi_\text{array}$ in~\eqref{EQ:Phi_Array}, which results in rotating the spectrum around the Rotman lens and hence the part of the spectrum falling into the port sector $[-\alpha_0,\alpha_0]$.
\begin{figure}[h!t]
    \centering
    \includegraphics[width=0.97\columnwidth]{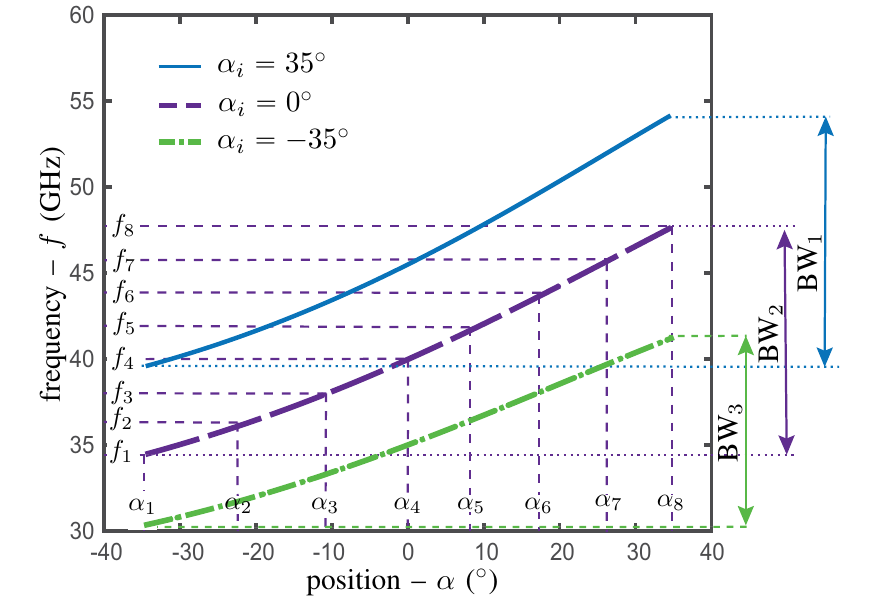}
        \caption{Frequency ($f$) versus port position ($\alpha$) of the lens RTSD in \figref{FIG:RTSD}, computed by~\eqref{EQ:Omega_In} for $\alpha_\text{i}=-35^\circ,0,35^\circ$, with $f_0=\omega_0/(2\pi)=40$~GHz and $N=2$.}
   \label{FIG:Deviation_alpha}
\end{figure}
\begin{figure}[h!t]
    \centering
    \includegraphics[width=0.8\columnwidth]{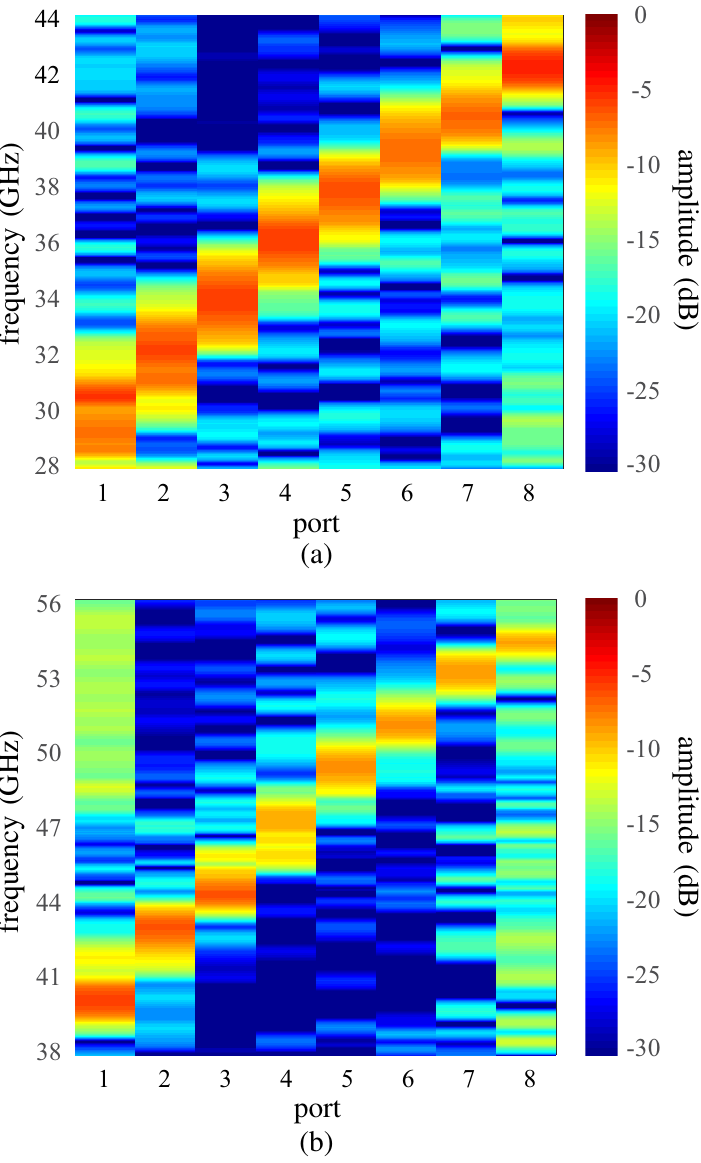}
        \caption{Full-wave simulated spectra versus lens port number in the RTSD of~\figref{FIG:Layout}(a) for different values of $\alpha_\text{i}$. (a)~$\alpha_\text{i}=-35^\circ$, or port~1 (bottom) in \figref{FIG:Layout}. (b)~$\alpha_\text{i}=35^\circ$ , or port~8 (top) in \figref{FIG:Layout}.}
   \label{FIG:S_Ports}
\end{figure}

The phase gradient term added by switching the input position is, according to the Rotman lens theory~\cite{JOUR:Rotman_Rotmanlens_1963}, $\omega\gamma d \sin{\alpha_\text{i}}/c$. Equation~\eqref{EQ:Phi_Array} generalizes then to
\begin{equation}\label{EQ:Phi_Array_In}
\Delta\phi_\text{array}(\omega)=2\pi N\frac{\omega}{\omega_0}-\frac{\omega}{c}\gamma d \sin{\alpha_\text{i}},
\end{equation}
which in turn generalizes Eq.~\eqref{EQ:Omega_Alpha} to
\begin{equation}\label{EQ:Omega_In}
\omega(\alpha)=\frac{2\pi N\omega_0c}{2\pi Nc-\omega_0d\gamma(\sin{\alpha}+\sin{\alpha_\text{i}})}.
\end{equation}
This relation confirms that increasing $\alpha_\text{i}$ from $-\pi/2$ to $\pi/2$ increases the frequency at any port, $\alpha_k$.

Figure~\ref{FIG:Deviation_alpha} plots the function $\omega(\alpha)$ given by~\eqref{EQ:Omega_In} for different values of $\alpha_\text{i}$. It confirms the predicted operation band tuning. Moreover, it shows the the bandwidth varies little with $\alpha_\text{i}$.

Figure~\ref{FIG:S_Ports} plots the full-wave simulated spectra versus input port position in the RTSD of \figref{FIG:Layout}(a) for $\alpha_\text{i}=-35^\circ$ and $\alpha_\text{i}=35^\circ$, complementing the results for $\alpha_\text{i}=0^\circ$ in \figref{FIG:Layout}(b).The operation bands and  bandwidths correspond to the theoretical predictions in \figref{FIG:Deviation_alpha}.

\begin{figure}[h!t]
    \centering
    \includegraphics[width=\columnwidth]{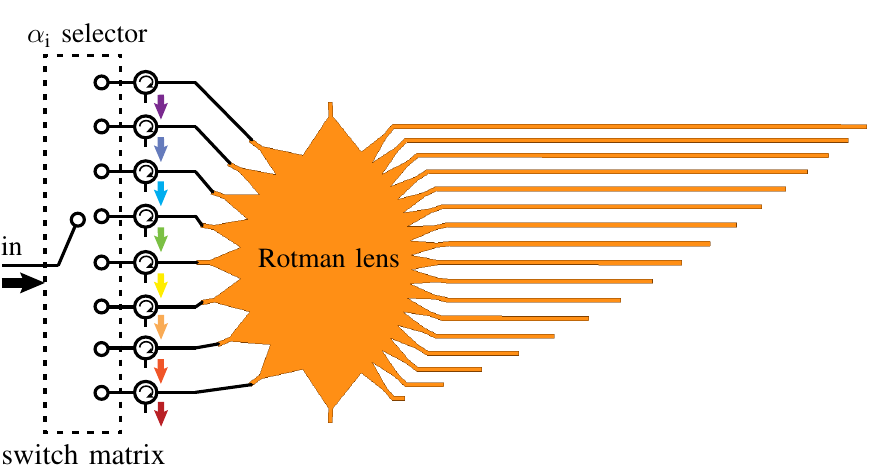}
        \caption{Implementation of operation band tuning RTSD-RTSS $\alpha_\text{i}$ using a switching matrix and circulators.}
   \label{FIG:Implement_2}
\end{figure}

Practically, the operation band tuning demonstrated in this section may be implemented using switching matrix and circulators, as shown in ~\figref{FIG:Implement_2}.

\section{Device Demonstration}\label{SEC:Demonstration}

\begin{figure}[h!t]
    \centering
    \includegraphics[width=\columnwidth]{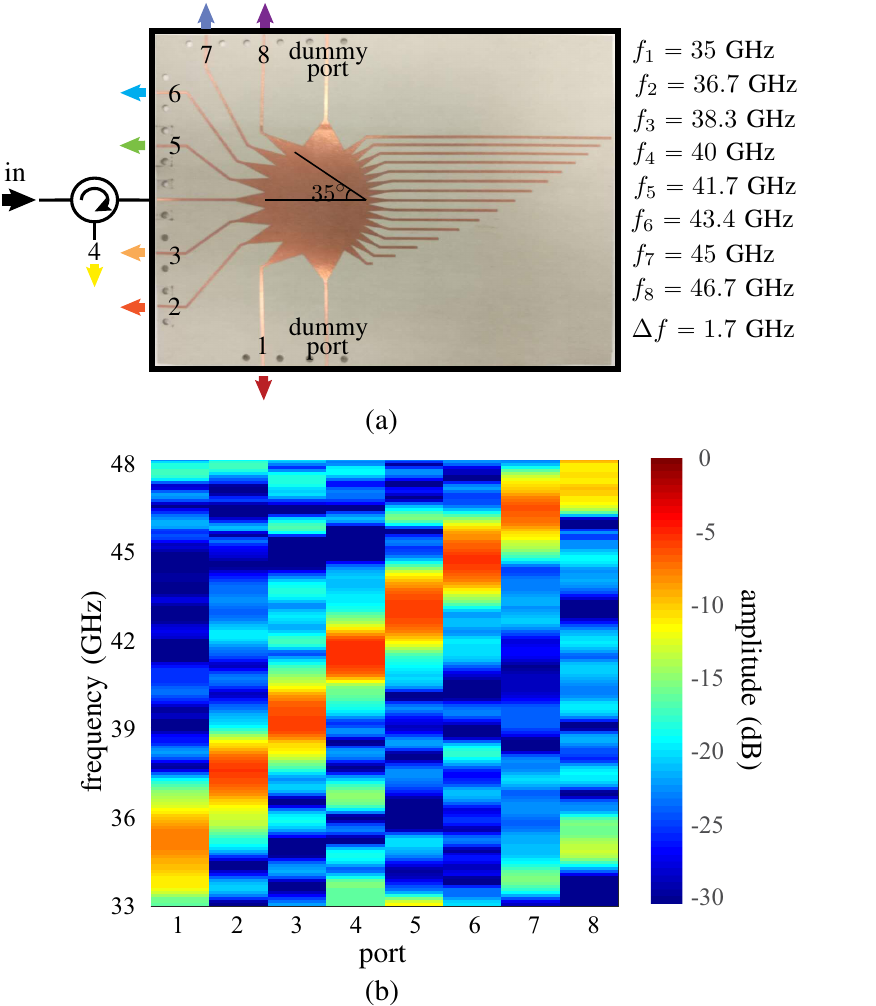}
        \caption{Spectral decomposition provided by the lens RTSD in \figref{FIG:Omega_Alpha} ($M=8$) with
        $N=2$ and uniform resolution. (a)~Fabricated prototype. (b)~Measured spectrum distribution versus lens port number.}
   \label{FIG:prototype}
\end{figure}

The RLSD of~\figref{FIG:Layout} was fabricated and measured. Figure~\ref{FIG:prototype}(a) shows the prototype while \figref{FIG:prototype}(b) plots the corresponding measured spectrum versus output port number. The measured spectrum distribution agrees very well with the theoretical prediction in \figref{FIG:Layout}(b) at least down to $-15$~dB.

\begin{figure}[h!t]
    \centering
    \includegraphics[width=0.8\columnwidth]{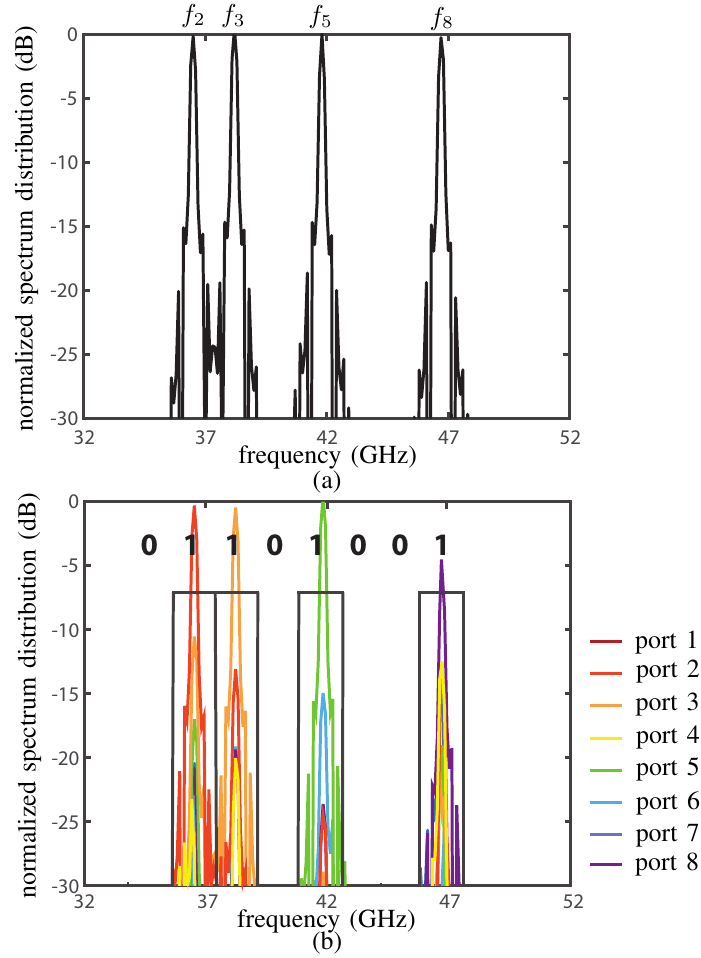}
        \caption{Spectra of the input and output signals in the RTSS (\figref{FIG:principle}) for the fabricated lens RTSD in \figref{FIG:prototype}. (a)~Multi-channel input signal with active channels $f_2=36.7$~GHz, $f_3=38.3$~GHz, $f_5=41.7$~GHz and $f_8=46.7$ GHz. (b)~Output signals just before and after the detectors in \figref{FIG:principle}.}
   \label{FIG:demonstration}
\end{figure}

Figure~\ref{FIG:demonstration} plots the spectra of the input and output signals of the RTSS (\figref{FIG:principle}) for the fabricated lens RTSD in \figref{FIG:prototype} with the 4 active channels $f_2=36.7,f_3=38.3,f_5=41.7,f_8=46.7$~GHz. The RTSS device, assuming a threshold between $-5$~dB and $-10$~dB properly detects these active channels, and hence informs on the availability of the channels $f_1=35,f_4=40,f_6=43.4,f_7=45$~GHz.

The implementation of the bandwidth-resolution tuning and operation band tuning versions of the RTSS demonstrated in Figs.~\ref{FIG:prototype} and \ref{FIG:demonstration} may be realized in the electronic architectures in \figref{FIG:Implement_1} and~\ref{FIG:Implement_2} has therefore not been implemented in the lab. Since these implementation would represent straightforward fabrication without added essential information, we have not implemented them. However, we have verified that the corresponding spectra of the fabricated prototype for all the tuning conditions in the operation band tuning RTSS and by full-wave simulation for the bandwidth-resolution tuning RTSS, and found, without surprise, excellent agreement in all cases.

\section{Conclusion}\label{SEC:Conclusion}

We have presented and demonstrated a Rotman-lens spectrum decomposer (RLSD) real-time spectrum-sniffer (RTSS) for cognitive radio. This system has superior features compared to previous RTSS devices. Moreover, it is easily amenable to bandwidth-resolution tuning and operation band tuning. the RLSD RTSS may thus find wide applications in future communication systems, particularly in the millimeter-wave frequency range where the device size is perfectly accommodatable.

\bibliographystyle{IEEEtran}
\bibliography{IEEEabrv,Xiaoyi_Reference}

\begin{thebibliography}{10}
\providecommand{\url}[1]{#1}
\csname url@samestyle\endcsname
\providecommand{\newblock}{\relax}
\providecommand{\bibinfo}[2]{#2}
\providecommand{\BIBentrySTDinterwordspacing}{\spaceskip=0pt\relax}
\providecommand{\BIBentryALTinterwordstretchfactor}{4}
\providecommand{\BIBentryALTinterwordspacing}{\spaceskip=\fontdimen2\font plus
\BIBentryALTinterwordstretchfactor\fontdimen3\font minus
  \fontdimen4\font\relax}
\providecommand{\BIBforeignlanguage}[2]{{%
\expandafter\ifx\csname l@#1\endcsname\relax
\typeout{** WARNING: IEEEtran.bst: No hyphenation pattern has been}%
\typeout{** loaded for the language `#1'. Using the pattern for}%
\typeout{** the default language instead.}%
\else
\language=\csname l@#1\endcsname
\fi
#2}}
\providecommand{\BIBdecl}{\relax}
\BIBdecl

\bibitem{Jour:Liang_2011_TVT_CognitiveRadio}
Y.~C. Liang, K.~C. Chen, G.~Y. Li, and P.~Mahonen, ``Cognitive radio networking
  and communications: an overview,'' \emph{IEEE Trans. Veh. Technol.}, vol.~60,
  no.~7, pp. 3386--3407, Sept. 2011.

\bibitem{Book:Osseiran_2016_5G}
A.~Osseiran, J.~F. Monserrat, and P.~Marsch, \emph{5G mobile and wireless
  communications technology}.\hskip 1em plus 0.5em minus 0.4em\relax Cambridge
  University Press, 2016.

\bibitem{Jour:2013_MwMag_Caloz}
C.~Caloz, S.~Gupta, Q.~Zhang, and B.~Nikfal, ``Analog signal processing: A
  possible alternative or complement to dominantly digital radio schemes,''
  \emph{IEEE Microw. Mag.}, vol.~14, no.~6, pp. 87 -- 103, Sept. 2013.

\bibitem{Jour:Zou_2017_TWC_DCMA}
L.~Zou, S.~Gupta, and C.~Caloz, ``Real-time dispersion code multiple access for
  high-speed wireless communications,'' \emph{IEEE Transactions on Wireless
  Communications}, vol.~17, no.~1, pp. 266--281, Jan 2018.

\bibitem{JOUR:2003_TMTT_Laso}
M.~A.~G. Laso, T.~Lopetegi, M.~J. Erro, D.~Benito, M.~J. Garde, M.~A. Muriel,
  M.~Sorolla, and M.~Guglielmi, ``Real-time spectrum analysis in microstrip
  technology,'' \emph{IEEE Trans. Microw. Theory Techn.}, vol.~51, no.~3, pp.
  705 -- 717, Mar. 2003.

\bibitem{JOUR:2015_Gomez-Tornero_TMTT_SIW_Multiplexer}
J.~L. G\'{o}mez-Tornero, A.~J. Mart\'{i}nez-Ros, S.~Mercader-Pellicer, and
  G.~Goussetis, ``Simple broadband quasi-optical spatial multiplexer in
  substrate integrated technology,'' \emph{IEEE Trans. Microw. Theory Techn.},
  vol.~63, no.~5, pp. 1609--1620, May 2015.

\bibitem{JOUR:2009_Gupta_TMTT_RTSA}
S.~Gupta, S.~Abielmona, and C.~Caloz, ``Microwave analog real-time spectrum
  analyzer ({RTSA}) based on the spectral-spatial decomposition property of
  leaky-wave structures,'' \emph{IEEE Trans. Microw. Theory Techn.}, vol.~57,
  no.~12, pp. 2989--2999, Dec. 2009.

\bibitem{JOUR:2014_MWCL_Nikfal}
B.~Nikfal, Q.~Zhang, and C.~Caloz, ``Enhanced-{SNR} impulse radio transceiver
  based on phasers,'' \emph{IEEE Microw. Wireless Compon. Lett.}, vol.~24,
  no.~11, pp. 778 -- 780, Nov. 2014.

\bibitem{Conf:Wang_AP-S2017_Phaser}
X.~Wang and C.~Caloz, ``Phaser-based polarization-dispersive antenna and
  application to encrypted communication,'' in \emph{Proc. IEEE Int. Symp. on
  Antennas Propag.}, July 2017.

\bibitem{JOUR:2011_TMTT_Nikfal}
B.~Nikfal, S.~Gupta, and C.~Caloz, ``Increased group delay slope loop system
  for enhanced-resolution analog signal processing,'' \emph{IEEE Trans. Microw.
  Theory Techn.}, vol.~59, no.~6, pp. 1622 -- 1628, Jun. 2011.

\bibitem{JOUR:2012_MWCL_Horii}
Y.~Horii, S.~Gupta, B.~Nikfal, and C.~Caloz, ``Multilayer broadside-coupled
  dispersive delay structures for analog signal processing,'' \emph{IEEE
  Microw. Wireless Compon. Lett.}, vol.~22, no.~1, pp. 1 -- 3, Jan. 2012.

\bibitem{JOUR:2012_TMTT_Zhang}
Q.~Zhang, S.~Gupta, and C.~Caloz, ``Synthesis of narrowband reflection-type
  phasers with arbitrary prescribed group delay,'' \emph{IEEE Trans. Microw.
  Theory Techn.}, vol.~60, no.~8, pp. 2394 -- 2402, Aug. 2012.

\bibitem{JOUR:2012_TMTT_Shulabh}
S.~Gupta, D.~L. Sounas, H.~V. Nguyen, Q.~Zhang, and C.~Caloz, ``{CRLH-CRLH}
  {C}-section dispersive delay structures with enhanced group-delay swing for
  higher analog signal processing resolution,'' \emph{IEEE Trans. Microw.
  Theory Techn.}, vol.~60, no.~12, pp. 3939 -- 3949, Dec. 2012.

\bibitem{JOUR:2014_JRMCAE_Zhang}
Q.~Zhang, S.~Gupta, and C.~Caloz, ``Synthesis of broadband phasers formed by
  commensurate {C}- and {D}-sections,'' \emph{Int. J. RF Microw. Comput. Aided
  Eng.}, vol.~24, no.~3, pp. 322 -- 331, May 2014.

\bibitem{JOUR:2015_TMTT_Gupta}
S.~Gupta, Q.~Zhang, L.~Zou, L.~Jiang, and C.~Caloz, ``Generalized coupled-line
  all-pass phasers,'' \emph{IEEE Trans. Microw. Theory Techn.}, vol.~63, no.~3,
  pp. 1 -- 12, Mar. 2015.

\bibitem{Conf:Wang_URSIGASS2017}
X.~Wang, L.~Zou, and C.~Caloz, ``Tunable c-section phaser for dynamic analog
  signal processing,'' in \emph{Proc. XXXIInd URSI GASS}, Aug 2017.

\bibitem{Jour:Nikfal_2012_MWCL_SpectrumSniffer}
B.~Nikfal, D.~Badiere, M.~Repeta, B.~Deforge, S.~Gupta, and C.~Caloz,
  ``Distortion-less real-time spectrum sniffing based on a stepped group-delay
  phaser,'' \emph{IEEE Microw. Wirel. Compon. Lett.}, vol.~22, no.~11, pp.
  601--603, 2012.

\bibitem{Conf:Fusco_Multiplerxer_2012}
Y.~Zhang and V.~Fusco, ``Rotman lens used as a demultiplexer/multiplexer,'' in
  \emph{2012 42nd European Microwave Conference}, Oct. 2012, pp. 164--167.

\bibitem{Jour:Wang_2017_RLSD}
X.~Wang, A.~Akbarzadeh, L.~Zou, and C.~Caloz, ``Flexible-resolution,
  arbitrary-input and tunable {R}otman lens spectrum decomposer ({RL-SD}),''
  \emph{arXiv preprint arXiv:1710.09226}, 2017.

\bibitem{JOUR:Rotman_Rotmanlens_1963}
W.~Rotman and R.~Turner, ``Wide-angle microwave lens for line source
  applications,'' \emph{IEEE Trans. Antennas Propag.}, vol.~11, no.~6, pp.
  623--632, Nov. 1963.

\bibitem{JOUR:Hansen_Rotmanlenses_1991}
R.~C. Hansen, ``Design trades for {R}otman lenses,'' \emph{IEEE Trans. Antennas
  Propag.}, vol.~39, no.~4, pp. 464--472, Apr. 1991.

\bibitem{Jour:Rotman_ProcIEEE_TrueTimeDelay}
R.~Rotman, M.~Tur, and L.~Yaron, ``True time delay in phased arrays,''
  \emph{Proc. IEEE}, vol. 104, no.~3, pp. 504--518, March 2016.

\bibitem{JOUR:2014_CJE_Vashist_ReviewRotamLens}
S.~Vashist, M.~K. Soni, and P.~K. Singhal, ``A review on the development of
  rotman lens antenna,'' \emph{Chinese Journal of Engineering}, 2014.

\bibitem{Thesis:Dong_RotmanLens_2009}
J.~Dong, ``Microwave lens designs: Optimization, fast simulation algorithms,
  and 360-degree scanning techniques,'' Ph.D. dissertation, Virginia Tech,
  2009.

\bibitem{Jour:1996_Smit_AWG}
M.~K. Smit and C.~V. Dam, ``Phasar-based {WDM}-devices: Principles, design and
  applications,'' \emph{IEEE J. Sel. Top. Quantum Electron.}, vol.~2, no.~2,
  pp. 236--250, Jun. 1996.

\bibitem{BK:2011_Pozar}
D.~M. Pozar, \emph{Microwave Engineering 4th Ed.}\hskip 1em plus 0.5em minus
  0.4em\relax John Wiley, 2011.

\end{thebibliography}
\end{document}